\documentstyle[prl,preprint,aps]{revtex}
\newcommand{\beq}{\begin{equation}}
\newcommand{\eeq}{\end{equation}}
\begin{document}
\baselineskip=18pt
\title{ Quasiparticle picture of high temperature superconductors
in the frame of Fermi liquid with the fermion condensate}

\author{M.Ya. Amusia$^{a,b}$ and V.R. Shaginyan$^{a,c}$
\footnote{E--mail: vrshag@thd.pnpi.spb.ru}}
\address{$^{a\,}$The
Racah Institute of Physics, the Hebrew University, Jerusalem 91904,
Israel;\\ $^{b\,}$Physical-Technical Institute, Russian Academy of
Sciences, 194021 St. Petersburg, Russia;\\ $^{c\,}$ Petersburg Nuclear
Physics Institute, Russian Academy of Sciences, Gatchina, 188350,
Russia}
\maketitle

\begin{abstract}
A model of a Fermi liquid with the fermion condensate (FC) is
applied to the consideration of quasiparticle excitations in high
temperature superconductors, in their superconducting and normal
states. Within our model the appearance of the fermion condensate
presents a quantum phase transition, that separates the regions of
normal and strongly correlated electron liquids.  Beyond the phase
transition point the quasiparticle system is divided into two
subsystems, one containing normal quasiparticles and the other ---
fermion condensate localized at the Fermi surface and
characterized by almost dispersionless single-particle
excitations. In the superconducting state the quasiparticle
dispersion in systems with FC can be presented by two straight
lines, characterized by effective masses $M^*_{FC}$ and $M^*_L$,
respectively, and intersecting near the binding energy which is of
the order of the superconducting gap. This same quasiparticle
picture persists in the normal state, thus manifesting itself over
a wide range of temperatures as new energy scales. Arguments are
presented that fermion systems with FC have features of a "quantum
protectorate" \cite{rlp,pa}.
\end{abstract}

\pacs{ PACS numbers: 71.27.+a, 74.20.Fg, 74.25.Jb}

\section {INTRODUCTION}

The single-particle excitations in ordinary Fermi liquids, e.g.,
electron liquid of metals, and their energy scales, define the
major part of their low temperature properties. High temperature
superconductors (HTS) are characterized by a number of striking
features. Among them are extremely high transition temperatures,
$T_c$, and the linear dependence of the resistivity on temperature
at $T>T_c$. The former behavior has been related to the existence
of the only one relevant energy scale, that is the temperature
$T$, which leads to a central conclusion of the Marginal Fermi
Liquid (MFL) that the one-particle self energy depends  only on
temperature and frequency, and not on momentum \cite{var}. Such a
behavior demonstrates that contributions from phonons excitations,
collective states, or impurities to the self energy are
inessential. All this permits to introduce the notion of a
"quantum protectorate", as a state of a system with so strong
correlations that these conventional effects are inessential
\cite{rlp,pa}. On the other hand, recent discovery of a new energy
scale for quasiparticle dispersion in superconducting and normal
states of Bi$_2$Sr$_2$CaCu$_2$O$_{8+\delta}$ \cite{blk,krc} can
bring new insight to physics of HTS, imposing serious constraints
upon possible theories of HTS. The newly discovered additional
energy scale manifests itself as a break in the quasiparticle
dispersion near $(50-70)\,$meV, which results in a drastic change
of the quasiparticle velocity \cite{blk,krc,vall}. Such a behavior
is qualitatively different from what one could expect in a normal
Fermi liquid. Moreover, this behavior can hardly be understood in
the frames of either MFL theory or the quantum protectorate since
there are no additional energy scales in these theories
\cite{pa,var}. One could suggest that this observed strong
self-energy effect, leading to the new energy scale, is due to the
electron coupling with collective excitations. But in that case
one has to give up the quantum protectorate idea, which would
contradict observations \cite{rlp,pa}.

The aim of our paper is to show that without giving up the quantum
protectorate idea the energy scale for quasiparticle dispersion
can be naturally explained within the model of correlated electron
liquid with FC. In Sec. 2, we review the general features of Fermi
systems with the FC, showing that an electron liquid of low
density  inevitably undergoes the fermion condensation quantum
phase transition (FCQPT). In Sec. 3 we consider the
superconducting state, which take place in the presence of the FC
and describe the quasiparticle dispersion and lineshape. Finally,
in Sec. 4, we summarize our main results.

\section {THE GENERAL FEATURES OF ELECTRON LIQUID WITH FC}

To describe a correlated electron liquid a conventional way can be
used, assuming that the correlated regime is connected to the
noninteracting Fermi gas by adiabatic continuity in the same way
as in the framework of the Landau normal Fermi liquid theory
\cite{lan}. But a question exists whether this is possible at all.
Most likely, the answer is negative. Therefore, we direct our
attention to a model, in the frame of which a strongly correlated
electron liquid is separated from conventional Fermi liquid by a
phase transition related to the onset of FC \cite{ks,vol}.

Let us start by considering the key points of the FC theory. FC is
a new solution of the Fermi liquid theory equations\cite{lan} for
the quasiparticle distribution function $n(p,T)$ \beq \frac{\delta
(F-\mu N)}{\delta
n(p,T)}=\varepsilon(p,T)-\mu(T)-T\ln\frac{1-n(p,T)}{n(p,T)}=0,\eeq
which depends on the momentum $p$ and temperature $T$. Here $F$ is
the free energy, and $\mu$ is the chemical potential, while
\beq\varepsilon(p,T)=\frac{\delta E[n(p)]}{\delta n(p,T)},\eeq is
the quasiparticle energy. This energy is a functional of $n(p,T)$
just like the total energy $E[n(p)]$ and the other thermodynamic
functions. Eq. (1) is usually presented as the Fermi-Dirac
distribution, \beq
n(p,T)=\left\{1+\exp\left[\frac{(\varepsilon(p,T)-\mu)}
{T}\right]\right\}^{-1}.\eeq At $T\to 0$ one gets from Eqs. (1),
(3) the standard solution $n_F(p,T\to0)\to\theta(p_F-p)$, with
$\varepsilon(p\simeq p_F)-\mu=p_F(p-p_F)/M^*_L$, where $p_F$ is
the Fermi momentum, and $M^*_L$ is the Landau effective mass
\cite{lan}, \beq \frac{1}{M^*_L}
=\frac{1}{p}\frac{d\varepsilon(p,T=0)}{dp}|_{p=p_F}.\eeq It is
implied that $M^*_L$ is positive and finite at the Fermi momentum
$p_F$. As a result, the $T$-dependent corrections to $M^*_L$, to
the quasiparticle energy $\varepsilon (p)$, and other quantities,
start with $T^2$-terms. But this solution is not the only one
possible. There exist ``anomalous" solutions of Eq. (1) associated
with the so-called fermion condensation \cite{ks,ksk}. Being
continuous and satisfying the inequality $0<n(p)<1$ within some
region in $p$, such solutions $n(p)$ admit a finite limit for the
logarithm in Eq. (1) at $T\rightarrow 0$ yielding, \beq
\varepsilon(p)=\frac{\delta E[n(p)]} {\delta n(p)} =\mu; \:
p_i\leq p \leq p_f. \eeq At $T=0$ Eq. (5) determines the FC
quantum phase transition (FCQPT), possessing solutions at some
$r_s=r_{FC}$ as soon as the effective inter-electron interaction
becomes sufficiently strong \cite{ksz}. In a simple electron
liquid, the effective inter-electron interaction is proportional to
the dimensionless average interparticle distance $r_s\sim
r_0/a_B$, with $r_0\sim 1/p_F$ being the average distance, and
$a_B$ is the Bohr radius. Equation (5) leads to the minimal value
of $E$ as a functional of $n(p)$ when in system under
consideration a strong rearrangement of the single particle
spectra can take place. We see from Eq. (5) that the occupation
numbers $n(p)$ become variational parameters: the solution $n(p)$
takes place if the energy $E$ can be lowered by alteration of the
occupation numbers. Thus, within the region $p_i<p<p_f$, the
solution $n(p)=n_F(p)+\delta n(p)$ deviates from the Fermi step
function $n_F(p)$ in such a way that the energy $\varepsilon(p)$
stays constant while outside this region $n(p)$ coincides with
$n_F(p)$. It follows from the above consideration that the
superconductivity order parameter $\kappa({\bf p})=\sqrt{n({\bf
p})(1-n({\bf p}))}$ has a nonzero value over the region occupied
by FC. The superconducting gap $\Delta({\bf p})$ being linear in
the coupling constant of the particle-particle interaction
$V_{pp}$ gives rise to the high value of $T_c$ because one has
$2T_c\simeq \Delta$ within the standard Bardeen-Cooper-Schrieffer
(BCS) theory \cite{dkss}. On the other hand, even at $T=0$,
$\Delta$ can vanish, provided $V_{pp}$ is repulsive or absent.
Then, as it is seen from Eq. (5), the Landau quasiparticle system
becomes separated into two subsystems. The first contains the
Landau quasiparticles, while the second, related to FC, is
localized at the Fermi surface and formed by dispersionless
quasiparticles. As a result, the standard Kohn-Sham scheme for the
single particle equations is no longer valid beyond the point of
the FC phase transition \cite{vsl}. Such a behavior of systems
with FC is clearly different from what one expects from the well
known local density calculations. Therefore these calculations are
hardly applicable to describe systems with FC. It is also seen
from Eq. (5) that a system with FC has a well-defined Fermi
surface.

Let us assume that FC has just taken place, that is $p_i\to p_f\to
p_F$, and the deviation $\delta n(p)$ is small. Expanding
functional $E[n(p)]$ in Taylor's series with respect to $\delta
n(p)$ and retaining the leading terms, one obtains from Eq. (5),
\beq \mu=\varepsilon({\bf p}) =\varepsilon_0({\bf p})+\int
F_L({\bf p},{\bf p}_1)\delta n({\bf p_1}) \frac{d{\bf
p}_1}{(2\pi)^2}; \: p_i\leq p \leq p_f,\eeq where $F_L({\bf
p},{\bf p}_1)=\delta^2 E/\delta n({\bf p})\delta n({\bf p}_1)$ is
the Landau interaction. Both the Landau interaction and the
single-particle energy $\varepsilon_0(p)$ are calculated at
$n(p)=n_F(p)$. It is seen from Eq. (6) that the FC quasiparticles
forms a collective state, since their energies are defined by the
macroscopical number of quasiparticles within the region
$p_i-p_f$, and vice versa. The shape of the spectra is not
effected by the Landau interaction, which, generally speaking,
depends on the system's properties, including the collective
states, impurities, etc. The only thing defined by the interaction
is the width of the region $p_i-p_f$, provided the interaction is
sufficiently strong to produce the FC phase transition at all.
Thus, we can conclude that the spectra related to FC are of
universal form, being dependent, as we will see below, mainly on
temperature $T$, if $T>T_c$, or on the superconducting gap at
$T<T_c$.

According to Eq. (1), the single-particle excitations within the
interval $p_i-p_f$ have at $T_c\leq T\ll T_f$ the shape
$\varepsilon(p,T)$ linear in T \cite{dkss,kcs}, which can be
simplified at the Fermi level, \beq
\varepsilon(p,T)-\mu(T)=T\ln\frac{1-n(p)}{n(p)} \simeq
T\frac{1-2n(p)}{n(p)}|_{p\simeq p_F}. \eeq $T_f$ is the
temperature, above which FC effects become insignificant
\cite{dkss}, \beq
\frac{T_f}{\varepsilon_F}\sim\frac{p_f^2-p_i^2}{2M\varepsilon_F}
\sim\frac{\Omega_{FC}}{\Omega_F}.\eeq Here $\Omega_{FC}$ is the FC
volume, $\varepsilon_F$ is the Fermi energy, and $\Omega_F$ is the
volume of the Fermi sphere. We note that at $T_c\leq T\ll T_f$ the
occupation numbers $n(p)$ are approximately independent of $T$,
being given by Eq. (5). One can imagine that at these temperatures
dispersionless plateau $\varepsilon(p)=\mu$ given by Eq. (5) is
slightly turned counter-clockwise about $\mu$. As a result, the
plateau is just a little tilted and rounded off at the end points.
According to Eq. (7) the effective mass $M^*_{FC}$ related to FC
is given by, \beq \frac{p_F}{M^*_{FC}}\simeq
\frac{4T}{p_f-p_i}.\eeq To obtain Eq. (9) an approximation for the
derivative $dn(p)/dp\simeq -1/(p_f-p_i)$ was used. Having in mind
that $p_f-p_i\ll p_F$, and using (8) and (9) the following
estimates for the effective mass $M^*_{FC}$ are obtained, \beq
\frac{M^*_{FC}}{M_0} \sim
\frac{N(0)}{N_0(0)}\sim\frac{T_f}{T}.\eeq Eqs. (9) and (10) show
the temperature dependence of $M^*_{FC}$. In (10) $M_0$ denotes
the bare electron mass, $N_0(0)$ is the density of states of
noninteracting electron gas, and $N(0)$ is the density of states
at the Fermi level. Multiplying both sides of Eq. (9) by $p_f-p_i$
we obtain the energy scale $E_0$ separating the slow dispersing
low energy part, related to the effective mass $M^*_{FC}$, from
the faster dispersing relatively high energy part, defined by the
effective mass $M^*_{L}$ \cite{ars}, \beq E_0\simeq 4T.\eeq It is
seen from Eq. (11) that the scale $E_0$ does not depend on the
condensate volume. The single particle excitations are defined
according to Eqs. (7) and (9) by the temperature and by $n(p)$,
given by Eq. (5). Thus, we are led to the conclusion that the
one-electron spectrum is negligible disturbed by thermal
excitations, impurities, etc, so that one observes the features of
the quantum protectorate.

It is seen from Eq. (5) that at the point of FC phase transition
$p_f\to p_i\to p_F$, $M^*_{FC}$ and the density of states, as it
follows from Eqs. (5), (10), tend to infinity. One can conclude
that at $T=0$ the beginning of the FC phase transition is
connected to the absolute growth of $M^*_{L}$. It is essential to
have in mind, that the onset of the charge density wave
instability in a many-electron system, such as electron liquid,
which takes place as soon as the effective inter electron constant
reaches its critical value $r_s=r_{cdw}$ \cite{sns}, is preceded
by the unlimited growth of the effective mass. Therefore, the FC
takes place before the onset of the charge density wave. Hence, at
$T=0$, when $r_s$ reaches its critical value $r_{FC}<r_{cdw}$, the
FCQPT inevitably takes place. Thus, the formation of the FC can be
thought as a general property of an electron liquid of low
density, rather then an uncommon and anomalous solution of Eq. (1)
\cite{ksz}. Beyond the phase transition into the FC the condensate
volume is proportional to $(r_s-r_{FC})$ as well as
$T_f/\varepsilon_F\sim (r_s-r_{FC})$ at least when
$(r_s-r_{FC})/r_{FC}\ll 1$. Note, that such a behavior is in
accordance with the general properties of second order phase
transitions. Therefore, we can accept a model relating systems
with FC to HTS compounds, assuming that the effective coupling
constant $r_s$ increases with decreasing doping, exceeding its
critical value $r_{FC}$ at the levels corresponding to optimal
doped samples.  We remark, that this critical value $r_{FC}$
corresponds to the $r_s$ values of slightly overdoped samples
\cite{ksz}. On the other hand, there exist charge density waves or
strong fluctuations of charge ordering in underdoped HTS
\cite{grun}. As the result, our quite natural model suggests that
both quantities, $T_f$ and condensate volume $\Omega_{FC}$,
increase with decrease of doping. Thus, these values are higher in
underdoped samples as compared to overdoped ones provided $r_s$
meets the mentioned above conditions. According to experimental
facts the large density of states at the Fermi level reaches its
maximum in the vicinity of the Hove singularities, that is around
the point $(\pi,0)$ of the Brillouin zone, or $\bar{M}$, in HTS
compounds. The density of states reaches its minimum value at the
intersection of the so called nodal direction of the Brillouin zone
with the Fermi surface (see, e.g., \cite{shen}). The FC sets in
around the van Hove singularities \cite{kcs}, causing, according to
Eqs. (9) and (10), large density of states and large value of the
difference $(p_f-p_i)$ at the point $\bar{M}$. Then, the volume
$\Omega_{FC}$ and difference $(p_f-p_i)$ start to depend on the point
of the Fermi surface, say, on the angle $\phi$ along the Fermi
surface, which we count from the point $\bar{M}$ to the point at
which the density of states reaches its minimum value. Nonetheless,
as it is seen from Eq. (11), $E_0$ remains constant, being
independent of the angle. This differs essentially from the case for
the effective mass $M^*_{FC}$, that can strongly depend upon the
angle via the difference $(p_f(\phi)-p_i(\phi))$, as it is seen from
Eq. (9). It is pertinent to note that outside the FC region the
single particle spectrum is negligible affected by the temperature,
being defined by $M^*_L$, (see Eq. (4)), however calculated at $p\leq
p_i$ instead of at $p=p_F$. Thus, we come to the conclusion that a
system with FC is characterized by two effective masses: $M^*_{FC}$
that is related to the single particle spectrum at lower energy
scale, and $M^*_L$ describing the spectrum at higher energy scale.
These two effective masses manifest itself as a break in the
quasiparticle dispersion, which can be approximated by two straight
lines intersecting at the energy $E_0$.  This break takes place at
temperatures $T_c\leq T\ll T_f$ in accordance with the experimental
findings \cite{blk}, and, as we will see, at $T\leq T_c$
corresponding to the experimental facts \cite{blk,krc}, when the
superconducting state is based on the FC state. As to the
quasiparticle formalism, it is applicable to this problem since the
width $\gamma$ of single particle excitations is not large compared
to their energy being proportional $\gamma\sim T$ at $T>T_c$
\cite{dkss}. The lineshape can be approximated by a simple Lorentzian
\cite{ars}, being in accordance with experimental data obtained from
scans at a constant binding energy \cite{vall}. Then, FC serves as a
stimulating source of new phase transitions which lift the
degeneration of the spectrum. For example, FC can generate the
spin density wave, or antiferromagnetic phase transition, thus
leading to a whole variety of the system's properties. Then, the
onset of the charge density wave is preceded by the FCQPT, and
both of these phases can coexist at the sufficiently low density
when $r_s\geq r_{cdw}$. The simple consideration presented above
explains extremely large variety of HTS properties. We have seen
above that the superconductivity is strongly aided by the FC,
because both of the phases are characterized by the same order
parameter. As a result, the superconductivity, removing the
spectrum degeneration, ``wins the competition" with the other
phase transitions up to the critical temperature $T_c$. We turn
now to a consideration of quasiparticle dispersions  at $T\leq
T_c$.

\section{QUASIPARTICLE DISPERSIONS at $T\leq T_c$}

Let us discuss the origin of two effective masses $M^*_L$ and
$M^*_{FC}$ in the superconducting state resulting in nontrivial
quasiparticle dispersion and in alteration of the quasiparticle
velocity. As we will see our results are in a reasonably good
agreement with experimental data \cite{blk,krc,vall}. To simplify
the discussion let us put $T=0$. The ground state energy $E_{gs}$
of a system in the superconducting state is given by the BSC
theory formula \beq E_{gs}[\kappa({\bf p})]=E[n({\bf p})]+
E_{sc}[\kappa({\bf p})], \eeq where the occupation numbers $n({\bf
p})$ are connected to the order parameter, \beq n({\bf
p})=v^2({\bf p});\,\,\, \kappa({\bf p})=v({\bf
p})\sqrt{(1-v^2({\bf p}))}.\eeq The second term $E_{sc}[\kappa_p]$
on the right hand side of Eq. (12) is defined by the
superconducting contribution which in the simplest case of the
weak coupling regime is of the form, \beq E_{sc}[\kappa_p]= \int
V_{pp}({\bf p}_1,{\bf p}_2)\kappa({\bf p}_1) \kappa^*({\bf p}_2)
\frac{d{\bf p}_1d{\bf p}_2}{(2\pi)^4}. \eeq Consider a
two-dimensional electron liquid on a simple square lattice which
is in the superconducting state with d-wave symmetry of the order
parameter $\kappa({\bf p})$. In such a case, the long-range
component in momentum space of particle-particle interaction
$V_{pp}$ is repulsive, and the short-range component is relatively
dominant and attractive at small momenta \cite{abr}. Then the
short-range component can be taken as the first approximation to
$V_{pp}(q)\simeq -V_2\delta(q)$. The FC arises near the Van Hove
singularities, causing, as it follows from Eq. (10), large density
of states at these points \cite{kcs}. Hence, the different regions
with the maximal value $\Delta_1$ of the gap $\Delta$ and the
maximal density states overlap slightly \cite{ksk,abr,ams}.
Varying $E_{gs}$ given by Eq. (12) with respect to $\kappa({\bf
p})$ one finds, \beq \varepsilon({\bf p})-\mu=\Delta({\bf
p})\frac{1-2v^2({\bf p})}{2\kappa({\bf p})}.\eeq Here
$\varepsilon({\bf p})$ is defined by Eq. (2), and, \beq
\Delta({\bf p})=-\int V_{pp} ({\bf p}, {\bf p}_1) \sqrt{n({\bf
p}_1)(1-n({\bf p}_1))} \frac{d{\bf p}_1}{4\pi^2}.\eeq A few
remarks are in order at this point.  If $V_2\to 0$, then
$\Delta({\bf p})\to 0$, and Eq. (15) reduces to the equation, \beq
\varepsilon({\bf p})-\mu=0,\: {\mathrm {if}}\,\,\, 0<n({\bf
p})<1;\: \kappa({\bf p})\neq 0, \eeq presenting FC solutions,
defined by Eq. (5) \cite{dkss,vsl}. Thus, we come to the
conclusion that the function $\kappa({\bf p})$ is defined by Eq.
(5). While corrections to this function due to the pairing
interaction, being small, are of the order of $V_{pp}/F_L$,
because interaction $V_{pp}$ is obviously weak as compared to the
Landau interaction $F_L$. We note again remarkable peculiarity of
the FC phase transition at $T=0$: this transition is related to
spontaneous breaking of gauge symmetry, when the superconductivity
order parameter $\kappa({\bf p})$ has a nonzero value over the
region occupied by the fermion condensate, while $\Delta({\bf p})$
vanishes provided $V_{pp}=0$ \cite{dkss,vsl}. We can conclude that
the transition temperature of the FC phase transition is zero
because it is proportional to the gap, as it must be in the
standard theory of superconductivity. Therefore, the FC phase
transition is a quantum phase transition, while, at temperatures
$T\ll T_f$, the properties of considered many-electron system,
such as its single particle spectra, occupation numbers and etc.
are strongly influenced by the ``shadow" of FCQPT as it is seen
from Eqs. (9-11).

If $V_{pp}$ is nonzero but small as compared to $F_L$ and
attractive, the gap $\Delta$ is given by Eq. (16), with $n({\bf
p})$ and $\kappa({\bf p})$ being {\it determined} by Eq. (5).
Therefore, as it is seen from Eq. (16), the gap is {\it linear} in
the coupling constant of the particle-particle interaction $V_2$,
which leads to high values of both $\Delta_1$ and $T_c$ \cite{ks}.
Taking into account the $\delta$-function shape of the attractive
component of $V_{pp}$, we have  from Eq. (16) simple estimations
for the maximum value of the gap: $2\Delta_1\simeq V_2$. Since the
order parameter $\kappa({\bf p})$ is defined by Eq. (5), that is
determined by the interaction $F_L$, the shape of the gap,
including the location of its nodes is robust being resistant to
scattering upon impurities. We can again conclude that such
features resemble a quantum protectorate. Generally speaking, the
state of the quantum protectorate is preserved by the FCQPT. As
soon as the coupling constant $V_2$ becomes finite (although
remains small), the plateau $\varepsilon({\bf p})-\mu=0$ is
slightly tilted and rounded off at the end points, that is the
effective mass $M^*_{FL}$ becomes finite. To calculate $M^*_{FL}$,
we differentiate the both parts of Eq. (15) with respect to the
momentum $p$ and obtain the following relations, \beq
\frac{p_F}{M^*_{FL}}\simeq \frac{\Delta_1}{4\kappa({\bf
p})}\frac{1}{p_f-p_i}\simeq\frac{2\Delta_1}{p_f-p_i}|_{p\simeq
p_F}. \eeq Deriving Eq. (18) we took into account that the gap
achieves its maximum value $\Delta_1$ at the Fermi level, and
$\kappa(p\simeq p_F)\simeq 1/2$. We use the above approximation
for the derivative $dn/dp$ and Eq. (13) to calculate the
derivative $d(v^2)/dp$: \beq \frac{d(v^2(p))}{dp}\simeq
-\frac{1}{(p_f-p_i)}.\eeq Now, one can conclude directly from Eq.
(18), that the following relation is valid \cite{ars} \beq
E_0\simeq\frac{(p_f-p_i)p_F}{M^*_{FC}}\simeq 2\Delta_1. \eeq It is
seen from Eq. (19) that again, this time at $T=0$, the
quasiparticle dispersion can be presented by two straight lines
characterized by two effective masses, $M^*_{FC}$ and $M^*_L$,
respectively, and intersecting near the binding energy
$E_0\simeq2\Delta_1$. Evaluations of the effective mass $M^*_{FL}$
at $T\to T_c$ is straightforward and similar to the presented
above. It is important that at finite temperatures we have to
replace Eq. (13) by  another equation of the BCS theory, namely by
\beq v^2({\bf p})=\frac{n({\bf p})-f({\bf p})}{1-2f({\bf p})},\eeq
where, \beq f({\bf p})=\frac{1}{1+\exp[E({\bf p})/T]};\,\,\,
E({\bf p})=\sqrt{(\varepsilon({\bf p})-\mu)^2+\Delta^2({\bf
p})}.\eeq After performing some straightforward algebraic
transformations and taking into account that the function $f({\bf
p})$ reaches its maximum at the Fermi level, while $E({\bf p})\ll
T$, we obtain instead of Eq. (19) the following equations \beq
\frac{d(v^2(p))}{d p}\simeq -\frac{1}{(p_f-p_i)(1-2f({\bf
p}))}\simeq -\frac{2T}{E(p)(p_f-p_i)}|_{T\to T_c}.\eeq Deriving
Eq. (23) we use the former approximation for $dn/dp$ and have in
mind that at $T\ll T_f$ the occupation numbers are temperature
independent and defined by Eq. (5). Differentiating Eq. (15) with
respect to the momentum $p$ and taking into account Eq. (23), we
estimate the effective mass as \beq M^*_{FL}\simeq
\frac{p_F(p_f-p_i)}{4T}.\eeq As the result, we obtain from Eq.
(24) an estimation for the energy scale, \beq E_0\simeq
\frac{(p_f-p_i)p_F}{M^*_{FC}}\simeq 4T.\eeq Comparing Eq. (20)
with (25) and bearing in mind that $2T_c\simeq \Delta_1$ we
conclude that both the effective mass $M^*_{FC}$ and the energy
scale $E_0$ are approximately temperature independent at $T\leq
T_c$, while Eqs. (24) and (25) match Eqs. (9) and (11) at $T=T_c$,
as one should expect.

The break separating the faster dispersing high energy part,
related to mass $M^*_L$, from the slower dispersing low energy
part defined by $M^*_{FC}$, is likely to be more pronounced in
underdoped samples. That is at least because of the rise of the
condensate volume $\Omega_{FC}$, leading to the growth of
$M^*_{FC}$ as it follows from Eqs. (9) and (18). We remind that
according our model the condensate volume $\Omega_{FC}$ is growing
with underdoping. It follows from Eqs. (9) and (18), that as one
moves along the Fermi surface from the nodal direction towards the
point $\bar{M}$, that is from minimal value of
$\Omega_{FC}(\phi)\sim (p_f^2(\phi)-p_i^2(\phi))$ towards the
maximal one, the ratio $M^*_{FC}/M^*_{L}$ grows in magnitude,
transforming the dispersion kink into a distinct break at the
point $\bar{M}$, or at the gap maximum $\Delta_1$ point. Thus, as
it follows from Eqs. (11) and (20) at $T\ll T_f$ there exists a
new energy scale defined by $E_0$, with $E_0\simeq 2\Delta_1$ at
$T\leq T_c$ and $E_0\simeq 4T$ at $T_c\leq T$. These results are
in good agreement with the experimental facts which show that at
$T\leq T_c$ as one moves towards $\bar{M}$ the dispersion kink
grows into the break separating the faster dispersing high energy
part of the single-particle spectrum from the slower dispersing
low energy part with a break in the slope near 50 meV \cite{blk},
or near 70 meV \cite{krc}. This effect is enhanced in underdoped
samples, and appears to persist at $T_c\leq T$ \cite{blk}.

Let us briefly discuss the lineshape of a quasiparticle peak
obtained from scans at a constant binding energy $\omega$
\cite{vall}, and at a constant momentum $q$, see e.g. \cite{shen}.
We recall that the lineshape $L$ of a quasiparticle peak can be
presented as a function of two variables: $L(q,\omega)$. Then, the
scans at constant binding energy is given by the function
$L(q,\omega=\omega_0)$, with $\omega_0$ is the binding energy of
the quasiparticle. Accordingly, $L(q=q_0,\omega)$ presents the
lineshape obtained from scans at the fixed momentum $q_0$
corresponding to the quasiparticle momentum. In order to consider
the width $\gamma$ of a quasiparticle peak, the special form of
the quasiparticle dispersion characterized by the two effective
masses should be taken into consideration. On the other hand,
scans at the constant energy reveal well defined single-particle
excitations with the width $\gamma\sim T$ at the Fermi level even
at the point $\bar{M}$ \cite{vall}. Considering $\gamma$ related
to the lineshape $L(q,\omega=\omega_0)$, provided $\omega_0\leq
E_0$, we can take into account only quasiparticles with the
effective mass $M^*_{FC}$, which can be large but finite. We can
do it because only quasiparticles with the energies less then
$E_0$ contribute to the width of a quasiparticle with the energy
$\omega_0$. Such a picture resembles the normal Fermi liquid
presented by quasiparticles with the effective mass $M^*_{FC}$.
The only difference is that now the effective mass, as it follows
from Eq. (9), depends on the temperature. As the result, we are
dealing with well-defined excitations of the width $\gamma\sim T$
\cite{dkss}, \beq \gamma\sim
\frac{(M^*_{FC})^3T^2}{\epsilon^2}\sim
\frac{(M^*_{FC})^3T^2}{(M^*_{FC})^2}\sim
T\frac{T_f}{\varepsilon_F}.\eeq Here $\epsilon$ is the dielectric
constant, which is proportional to the effective mass $M^*_{FC}$,
the latter being inversely proportional to $T$, see Eq. (9). This
result is in good agreement with the experimental findings cited
above \cite{vall}. Dealing with scans at constant $q$, which
correspond to the lineshape function $L(q=q_0,\omega)$, we have to
consider the contribution coming from quasiparticles with the mass
$M^*_L$ as well, because now there are no limits on the energy of
quasiparticles contributing to the width $\gamma$. In view of the
fact that the contribution of these excitations is enhanced by the
presence of FC, and these excitations start to contribute to the
lineshape at energies $\omega \geq E_0$, one can conclude that the
peak inevitably has a broadening which can hardly be interpreted
as standard width obtained from scans at a constant binding
energy. On the other hand, one may follow the procedure suggested
in \cite{krc}, using the Kramers-Kr\"{o}nig transformation to
construct the imaginary part of the self-energy starting with the
real one. As a result, the lineshape $L(q=q_0,\omega)$ of the
quasiparticle peak as a function of the binding energy $\omega$
possesses a complex peak-dip-hump structure \cite{krc} directly
defined by the existence of the two effective masses $M^*_{FC}$
and $M^*_L$ \cite{ars}.

\section{Concluding remarks}

We have discussed the model of a strongly correlated electron
liquid based on the FCQPT and applied it to high-temperature
superconductors. The FCQPT plays the role of a boundary separating
the region of a strongly interacting normal electron liquid from
the region of a strongly correlated electron liquid. It is
important to have in mind, that the onset of the charge density
wave instability in a many-electron system, which takes place as
soon as the effective inter-electron constant reaches its critical
value $r_s=r_{cdw}$, is preceded by the FCQPT. Hence at $T=0$,
when $r_s$ reaches its critical value $r_{FC}<r_{cdw}$, the FCQPT
inevitably takes place. Thus, the FC can be thought as a general
property of an electron liquid of the low density rather then a
unique phenomenon. We have shown that the quasiparticle dispersion
in systems with FC can be represented by two straight lines
characterized by the respective effective masses $M^*_{FC}$ and
$M^*_L$. At $T<T_c$, these lines intersect near the point $E_0\sim
2\Delta_1$, while above $T_c$, we have $E_0\sim 4T$. It is argued
that this strong change of the quasiparticle dispersion at $E_0$
can be enhanced in underdoped samples because of strengthening the
FC influence. The single-particle excitations and their width
$\gamma$ are also studied. Well-defined excitations with
$\gamma\sim T$ exist at the Fermi level even in the normal state.
This result is in line with the experimental findings determined
from the scans at constant binding energies $\omega$. We have
discussed also the lineshape obtained from scans at a constant
momentum $q$. In this case, the special form of the quasiparticle
dispersion should be taken into consideration. As the result, the
lineshape of the quasiparticle peak as a function of the binding
energy $\omega$ possesses a complex peak-dip-hump structure
directly defined by the existence of the two effective masses
$M^*_{FC}$ and $M^*_L$. We have also presented arguments that
fermion systems with FC have features of the quantum protectorate,
being separated from the normal Fermi liquid by the FC quantum
phase transition.

\section{ Acknowledgement}

V.R.S. thanks the Racah Institute of Physics at the Hebrew
University of Jerusalem where part of this work has been done for
hospitality. This research was supported in part by the Russian
Foundation for Basic Research under Grant No. 01-02-17189.

\end{document}